\documentclass[12pt]{iopart}

\usepackage{graphicx}
\usepackage{multirow}

\expandafter\let\csname equation*\endcsname\relax
\expandafter\let\csname endequation*\endcsname\relax
\usepackage{amsmath}
\usepackage{amssymb}

\usepackage{booktabs} 

\usepackage[lined,ruled,vlined]{algorithm2e}
\usepackage{algorithmicx}
\usepackage{algpseudocode}

\begin{document}

\title[Neural network replacing spectrum method]{Improving the Deconvolution of Spectra at Finite Temperatures by Replacing Spectrum with a Neural Network}

\author{Haidong Xie$^1$ \& Xueshuang Xiang$^2$ \& Yuanqing Chen$^1$ }
\address{$^1$China Academy of Aerospace Science and Innovation, Beijing 100082, China}
\address{$^2$Qian Xuesen Laboratory of Space Technology, China Academy of Space Technology, Beijing 100080, China}
\ead{xiehaidong@aliyun.com; xiangxueshuang@qxslab.cn}

\submitto{\JPCM}


\begin{abstract}
In condensed matter physics studies, spectral information plays an important role in understanding the composition of materials. However, it is difficult to obtain a material's spectrum information directly through experiments or simulations. For example, the spectral information deconvoluted by scanning tunneling spectroscopy suffers from the temperature broadening effect, which is a known ill-posed problem and makes the deconvolution results unstable. Existing methods, such as the maximum entropy method, tend to select an appropriate regularization to suppress unstable oscillations. However, the choice of regularization is difficult, and oscillations are not completely eliminated. We believe that the possible improvement direction is to pay different attention to different intervals. Combining stochastic optimization and deep learning, in this paper, we introduce a neural network-based strategy to solve the deconvolution problem. Because the neural network can represent any nonuniform piecewise linear function, our method replaces the target spectrum with a neural network and can find a better approximation solution through an accurate and efficient optimization. Experiments on theoretical datasets using superconductors demonstrate that the superconducting gap is more accurately estimated and oscillates less. Plug in real experimental data, our approach obtains clearer results for material analysis.
\end{abstract}

%
%
%
%
%

\section{Introduction}

In condensed matter physics, spectral analysis based on the density of states is one of the key ways to understand the composition of a material's properties~\cite{1983Condensed}.
For example, the widely recognized properties of metals and insulators depend on the spectral information in band theory~\cite{1998Metal}.
Of course, it also includes a series of new condensed matter materials represented by high-temperature superconductors, which are also inseparable from the composition analysis system based on spectra~\cite{1993Correlated}.
Whenever a new material is discovered or prepared, researchers are always eager to know the superconductivity gap of the material in the hope of testing the reliability of the theoretical model. However, it is not easy to accurately measure a material's superconductivity.

In fact, there is no experimental method to directly measure the spectrum.
At present, one feasible experimental strategy is to obtain the tunneling current at low temperature by using a scanning tunneling microscope (STM)~\cite{Park1987Scanning} and then deconvolute the zero temperature density of states according to the scanning tunneling spectrum.
This strategy presents difficulties because the process of solving zero temperature by finite temperature is a typical ill-posed problem in mathematics~\cite{PhysRevLett.6.57}.
Because temperature erases the high-frequency details of the spectrum information, the obtained zero-temperature information is unreliable.
Reducing the experimental temperature is a mathematically feasible method to alleviate the ill-posed problem, but it obviously makes the experiment difficult and unrealistic.
Regardless of the numerical precision of the input data and calculation process, the reliability of the output result is difficult to guarantee. Especially when our goal is to find the superconducting gap where the spectrum diverges.

To analyze the details of this process, we engage in a quantitative exploration. Fortunately, the theory of tunneling current ($I(\epsilon)$) at limited low temperature ($T$) based on the zero temperature density of states (DOS, $\rho(\omega)$) of materials is clearly understood, and this process can be expressed as the following formula in Eq.\eqref{eq:stm}~\cite{PhysRevLett.6.57},
\begin{eqnarray}
I(\epsilon) = (\frac{4 \pi e}{\hbar})  \int^{\infty}_{-\infty} 
\mathrm{d}\omega && |M|^2 \rho(\omega) \rho_{p}(\omega+\epsilon) \nonumber\\
&& \times[f(\omega)-f(\omega+\epsilon)],
\label{eq:stm}
\end{eqnarray}
where $\rho_{p}(\omega+\epsilon)$ is the DOS of the probe, $M$ is the tunneling matrix element, $f(\omega)={1}/{(e^{\beta \omega} + 1)}$ is the Fermi distribution function dependent on $\beta = 1/ k_B T$, and $k_B$, $e$ and $\hbar$ represent the Boltzmann constant, charge quantity and reduced Planck constant, respectively.

According to the theoretical formula of the temperature broadening effect from the Bardeen approximate formula as Eq.\eqref{eq:stm}, 
and to obtain the zero-temperature spectrum, the differential conductance must be deconvolved at a finite temperature, as shown in Eq.\eqref{eq:stm1}, with $M$ and $\rho_{p}$ treated as constants and eliminating $f(\omega)$, we have
\begin{eqnarray}
\sigma\left(\epsilon\right) &=& (\frac{\pi e^{2}\left\vert M \right\vert^{2}\rho_{p}\beta}{\hbar}) \nonumber\\
&& \times \int \mathrm{d} \omega\rho\left(\omega\right)\cosh^{-2}\frac{\beta\left(\omega+\epsilon\right)}{2}.
\label{eq:stm1}
\end{eqnarray}
This deconvolution problem of $\rho(\omega)$ from $\sigma(\epsilon)$ is essentially an ill-posed integral equation and can be transformed into a form that solves the matrix equation by discretization. The condition number of this problem is very large; thus, the result is unstable because of the sawtooth noise of input data $\sigma\left(\epsilon\right)$ and cannot be solved directly by least-squares fitting.

If the observed material is a superconductor, the DOS information not only accurately determines whether it enters the superconducting state but also estimates the superconductivity gap.
Therefore, a series of physical situations, with the above problems as examples, have great significance for scientific analysis and research.
In practical research, we often use the available data ($G(m)$) to deduce the dynamical information ($A(\omega)$), as in the general form in Eq.\eqref{eq:gen}~\cite{Teodorescu2013Mathematical,0Inverse},
\begin{eqnarray}
G(m) = \int^{\infty}_{-\infty} \mathrm{d} \omega K(m,\omega) A(\omega), m = 1,\dots,N, 
\label{eq:gen}
\end{eqnarray}
where $K(m,\omega)$ is a known kernel, and data $G(m)$ includes $N$ discrete values in total.
This process can usually be mathematically divided into the Fredholm integral equation of the first kind, which is a typical ill-posed problem in mathematics~\cite{Linear1999Rainer}.

Traditional methods are mainly based on easing the number of conditions~\cite{Jarrell2012MEM,Mishchenko2012Stochastic}, which include the pseudoinverse matrix method~(Pinv)~\cite{Golub1965Calculating} that ignores the minimal eigenvalues, the Tikhonov-Phillips regularization method~(TPRM)~\cite{Tikhonoff1943,Tikhonoff1963,Phillips1962} that uses regularization to increase small eigenvalues, and the maximum entropy method~(MEM)~\cite{JARRELL1996133,Vafayi2007Analytical} that regularizes the problem based on a baseline model. 
Based on regularizations, these methods alleviate the instability of the problem but still have oscillations and unclear boundaries. 
We think the reason for the oscillations and unclear boundaries is the dense uniform sampling needed to ensure the approximation.
The artificial neural network method~(ANN)~\cite{Fournier_2020,Yoon_2018,Arsenault_2017}, based on the idea of supervised learning in deep learning, uses a large number of $\rho(\omega)$ and $\sigma(\epsilon)$ correlation data for neural network model training.
As a result of the large-scale training data, this method can obtain accurate and fast results, even better and faster than MEM.
However, because of overreliance on training data, once the data sampling changes or the physical model changes, the performance cannot be guaranteed.
The stochastic optimization method~(SOM)~\cite{Skilling2002Probabilistic,Sandvik1998Stochastic,Mishchenko2000Diagrammatic} introduces dynamic and nonuniform discrete sampling to solve the problems explored in this paper. However, it is difficult to optimize the discrete mode in practice.
With a small number of samples, a model easily converges, but the accuracy of the results is limited. Using a large number of samples, the difficulty of convergence increases rapidly, and the results easily fluctuate.

To reduce the above difficulty, on the basis of dynamic optimization of nonuniform piecewise linear, we introduce a neural network into this problem and name it \textbf{neural network replacing spectrum (NNRS) method}.
The core idea is to use a neural network to address the objective function $\rho(\omega)$, and the training process of the neural network is used to replace the existing deconvolution algorithms. 
A fully connected neural network with a ReLU activation function is mathematically equivalent to a piecewise linear function; the training process makes it possible to adjust an appropriate nonuniform sampling distribution.
Considering the strong expression abilities of neural networks~\cite{Cybenko1989Approximation} and the priority of addressing low-frequency signals~\cite{Xu2018Understanding}, the experimental results show that, to some extent, our method can eliminate oscillation errors and improve the accuracy of the results.
Comparing the results of our NNRS with existing methods using the theoretical dataset and experimental data shows that our method has strong practical value for deconvoluting spectral problems.

The core characteristics of this method are summarized as follows:
\begin{enumerate} 
    \item Weaker oscillation: Using a neural network to learn the characteristics of low-frequency characteristic priority, the learned spectrum is smoother, and the oscillation is weaker.
    \item More accurate peak energy: Because the neural network can approximate a nonuniform piecewise linear function, the fitting effect can be improved by dense piecewise for the region with drastic spectral changes, and the peak energy estimation is more accurate.
    \item Nonnegative value at the energy gap: The ReLU activation function of the neural network ensures that the spectrum is nonnegative and maintains the efficiency and accuracy of training.
\end{enumerate}

\section{Related Work}\label{sec:2}


\subsection{Pseudo inverse matrix method}

The ill-posed problems in this paper can be generalized by Eq.\eqref{eq:gen}; for the STM problems explored in this paper, the formula is expressed as the tunneling current from Eq.\eqref{eq:stm} or as a differential conductance from Eq.\eqref{eq:stm1}.
First, the integral of Eq.\eqref{eq:stm1} needs to be discretized; considering that $\sigma(\epsilon)$ is usually discrete, the formula can be transformed into a matrix equation form.
\begin{eqnarray}
\vec{\sigma} = a \bar{A} \vec{\rho},
\label{eq:matrix}
\end{eqnarray}
where $\vec{\sigma}$ and $\vec{\rho}$ are vectors of variables $\epsilon$ and $\omega$, respectively, matrix $\bar{A}(\epsilon, \omega) \!=\! \cosh^{-2}({\beta\left(\omega\!+\!\epsilon\right)}/{2}) \text{dis}(\omega)$, with $\text{dis}(\omega)$ being the distance between adjacent discrete samples, and $a = ({\pi e^{2}\left\vert M \right\vert^{2}\rho_{p}\beta})/{\hbar}$.
Consequently, the calculation from the spectral information $\rho(\omega)$ to the measured data $\sigma(\epsilon)$ is good and can be simulated with high precision, but the process from $\sigma(\epsilon)$ to $\rho(\omega)$ is ill-posed. Because the condition number of matrix $\bar{A}$ is large, the result $\rho(\omega)$ is unstable due to the sawtooth noise of the input data $\sigma(\epsilon)$.
It can be summarized as the following loss function minimization with $\ell_2$ norm. These kinds of ill-posed problems cannot be solved directly by the least-squares fit methods.
\begin{eqnarray}
\min_\rho || \bar{A} \vec{\rho} - \vec{\sigma} ||_2^2 \qquad \text{or}\qquad  \vec{\rho} = \bar{A}^{-1} \vec{\sigma}.
\label{eq:loss}
\end{eqnarray}

Here, Pinv~\cite{Golub1965Calculating} is used to replace the inverse of the matrix, which avoids the restriction that the inverse matrix must require a square matrix, so the spectral information can be obtained.
Furthermore, given the very small eigenvalue of the matrix, because it is close to the machine error, the accuracy cannot be guaranteed, so it is not included in the optimization process to avoid divergence problems.

As a classical linear algebra method, Pinv appears more in textbooks than in practical applications.
The Pinv method has been integrated into most linear algebra libraries, such as scipy~\cite{2020SciPy-NMeth} in python: \textit{scipy.linalg.pinv}.

\subsection{Tikhonov-Phillips regularization method}

Considering the data errors, Eq.\eqref{eq:matrix} can solve the pseudoinverse matrix~\cite{Golub1965Calculating} by discarding eigenvalues of less than a given threshold. However, only in the case of low temperatures should it be reluctantly accepted; when the temperature rises, there will be violent oscillations because of the sawtooth noise of the input data, especially when it nears singularity.

The first effective way to deal with ill-posed problems was with TPRM, named for the first application of the idea to ill-posed integral equations by Tikhonov and Phillips~\cite{Tikhonoff1943,Tikhonoff1963,Phillips1962}; it is formalized as
\begin{eqnarray}
\min_\rho || \bar{A} \vec{\rho} - \vec{\sigma} ||^2_2 + \lambda || \bar{\Gamma} \vec{\rho} ||^2_2.
\end{eqnarray}
The TPRM method introduces the regularization constraint $\bar{\Gamma}$ on $\vec{\rho}$.
The introduction of this trick is equivalent to synchronously increasing the values of all eigenvalues of a matrix $\bar{A}$, e.g., $\bar{A} \rightarrow \bar{A} + \lambda \bar{I} $.
As a result, the condition number of the problem is improved, but the solution will deviate.

The main difficulty of this method is the selection of regularization constraints and correlation coefficients $\lambda$, which requires considerable human effort to optimize. Even so, the results are suppressed with large derivatives by regularization, especially when the spectral function has sharp edges or narrow peaks.

The TPRM is widely used in engineering, and it is relatively simple to implement the TPRM in the algorithm. It only needs to change the loss function from $||\bar{A} \vec{\rho} - \vec{\sigma}||$ to $||\bar{A} \vec{\rho}-\vec{\sigma}||^2 + \lambda ||\bar{\Gamma} \vec{\rho}||^2$ at the optimization objective level.

\subsection{Maximum entropy method}

To address the shortcomings of TPRM, MEM~\cite{JARRELL1996133,Vafayi2007Analytical} searches for the most likely solution $\vec{\rho}$ among the variational space by assuming the prior knowledge that $\vec{\rho}$ is close to a predefined function $D(\omega)$, called the default model. 
The regularization constraints are replaced by entropy,
\begin{eqnarray}
S[\vec{\rho}] = \int \mathrm{d} \omega \vec{\rho}(\omega) \ln (\frac{\vec{\rho}(\omega)}{D(\omega)}).
\end{eqnarray}
Entropy characterizes the deviation of $\vec{\rho}(\omega)$ from the default model $D(\omega)$.
If a large amount of information is known for $\vec{\rho}(\omega)$, a good default model $D(\omega)$ can be defined, and then MEM outperforms TPRM. However, the method highly relies on the default model, which is a serious drawback if the interrogated features of the spectra are very sensitive to the chosen default model~\cite{Jarrell2012MEM}.

Presently, MEM can be said to be one of the most widely used methods. A large number of similar problems have achieved good results by using MEM, for example, its use for real-valued single-orbital problems~\cite{PhysRevB.44.6011}.
Additionally, many studies have investigated the properties and optimization improvements of MEMs~\cite{PhysRevB.61.5147,1989Classic,PhysRevB.96.155128}.

To date, a large number of libraries have utilized the algorithmic functions of MEM, such as Maxent~\cite{Levy_2017}; and the code available in \textit{{https://github.com/TRIQS/maxent}}.

\subsection{Artificial neural network}

\begin{figure*}
  \centering
  \includegraphics[width=0.8\linewidth]{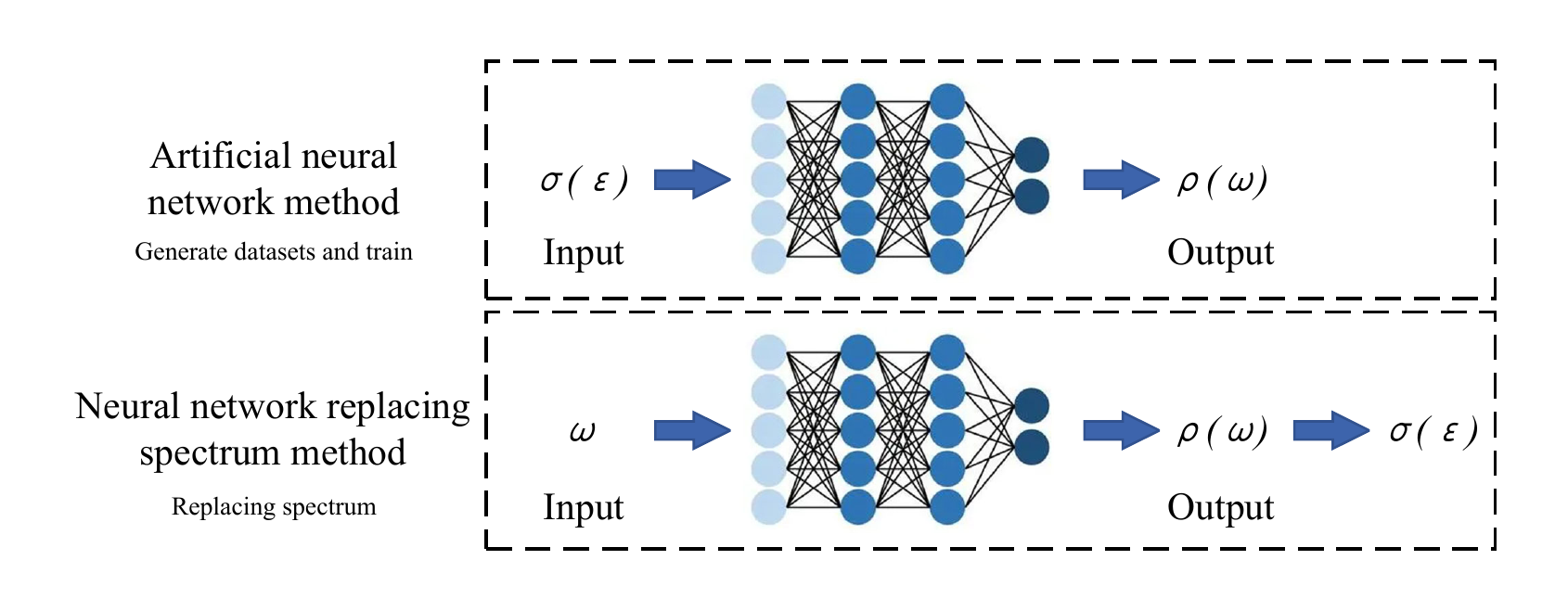}
  \caption{Comparison of the core ideas of the artificial neural network method and neural network replacing spectrum method. At the same time, the ANN method requires a preset dataset, but our NNRS method does not require this step.}
  \label{fig:ann}
\end{figure*}

With the development of deep learning, using neural networks to solve this problem has shown great potential.
Reference~\cite{Fournier_2020,Yoon_2018,Arsenault_2017} introduced an ANN based on supervised learning, with the same level of accuracy as the MEM, and the computational cost was reduced by almost three orders of magnitude.
The advantage of this method is that the process from $\vec{\rho}(\omega)$ to $\vec{\sigma}(\epsilon)$ corresponds to an errorless matrix operation, so a sufficiently large dataset can be generated for training.
\begin{eqnarray}
\vec{\sigma}(\epsilon) \rightarrow \text{ANN} \rightarrow \vec{\rho}(\omega).
\end{eqnarray}
This kind of algorithm has fast computing speeds; that is, once the neural network training is completed, the subsequent application only needs a single network inference process without any iterations; consequently, its speed for new data inputs is very fast.
In fact, for the determined physical model, increasing the training set size is a completely feasible and effective method to improve the accuracy because the equivalent parameter degrees of freedom in the actual physical model are usually low, and the neural network can completely learn the physical laws.

However, because it requires training steps, the training time is relatively long.
To ensure sufficient accuracy, training requires more than $10^5$ training samples. Such a large number of training samples may not be easy to obtain. Even if the model is well trained, the generalization ability of the model is inadequate~\cite{keskar2017largebatch}, which means that the model is dependent on the training data. If not enough cases are included in the training set, the trained model is likely to fail, including the calculation of new sampling data points and the calculation of new physical models.
Figure~\ref{fig:ann} shows the comparison of the ANN and NNRS methods. In the next section, we will discuss the difference in the calculation effect and mechanism of ANN compared with our method in detail.

Although this kind of method has great application potential, the academic community needs to jointly improve the training dataset and the training model base. Only when the model base has a sufficient scale can this kind of method have an ideal effect.
Considering that the training cost for large amounts of data is very high, the numerical experiment part of this paper does not include a comparison of ANN methods.

\subsection{Stochastic optimization method}

Compared with the previous methods, based on a large amount of prior knowledge of the model and uniformly fixed discrete sampling,
SOM~\cite{Skilling2002Probabilistic,Sandvik1998Stochastic,Mishchenko2000Diagrammatic} does not use any default model or impose any smoothing and only restricts prior knowledge to the normalization and positivity of the solution.
SOM uses a likelihood functional
\begin{eqnarray}
\vec{\rho} = \int \mathrm{d} \tilde{\rho}  \, \tilde{\rho} \, P(\tilde{\rho} | \vec{\sigma}),
\end{eqnarray}
where $\vec{\rho}$ is obtained as an average of particular solutions $\tilde{\rho}$ with the weight of likelihood function $P(\tilde{\rho} | \vec{\sigma})$. $P(\tilde{\rho} | \vec{\sigma})$ describes the corresponding probability by residual $|| \bar{A} \vec{\rho} - \vec{\sigma} ||$, although $\tilde{\rho}$ with a very small residual overfits the data $\vec{\sigma}$ with sawtooth noise; in practice, the sawtooth noise can be self-averaging in a sum over a large enough number of particular solutions if the residual is not kept too restrictive, which sets up an implicit regularization procedure. In some cases, if the residual can be ensured to be less than the given error threshold, then the influence of the weight can be ignored, and the above formula becomes a simple algebraic average of $N$ samples:
\begin{eqnarray}
\vec{\rho} = \frac{1}{N} \sum \tilde{\rho}.
\end{eqnarray}

Although SOM is less dependent on prior knowledge, the calculation process of a large number of particular solutions has many difficulties. Obtaining efficient independent solutions and ensuring the ergodicity of solutions are difficult to strictly guarantee.
Therefore, it is still difficult to achieve satisfactory results with this method~\cite{Mishchenko2012Stochastic}.

The SOM method may be a widely used solution second only to the MEM method. You can easily find the TRIQS/SOM library in GitHub to apply to computing projects~\cite{Krivenko_2019}. The software library contains a large number of commonly used physical models, which are easy to use. At the same time, many other scholars have also developed software packages with different implementation paths, which are also available on GitHub.

\section{Methodology}\label{sec:method}

In this section, we first discuss the possible advantages of nonuniform piecewise linear from the perspective of a piecewise strategy and then propose our NNRS method based on the expression ability of a neural network. We need to note that ANN is based on the idea of supervised learning, while our method adopts the idea of function approximations without a large amount of training data. The two methods are essentially different. 

\subsection{Condition number versus discrete sample rate}

\begin{figure}
  \centering
  \includegraphics[width=0.32\textwidth]{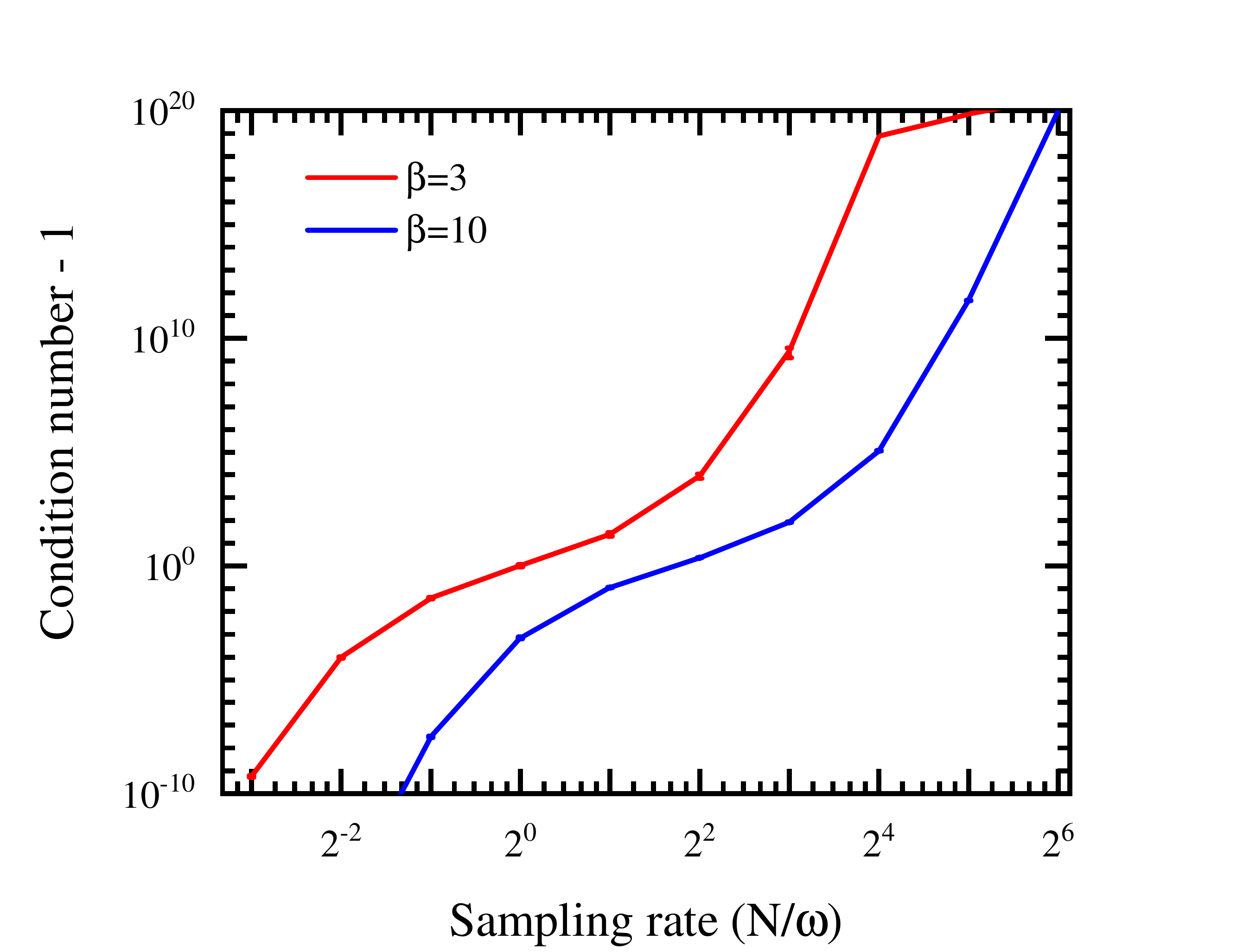}
  \includegraphics[width=0.32\textwidth]{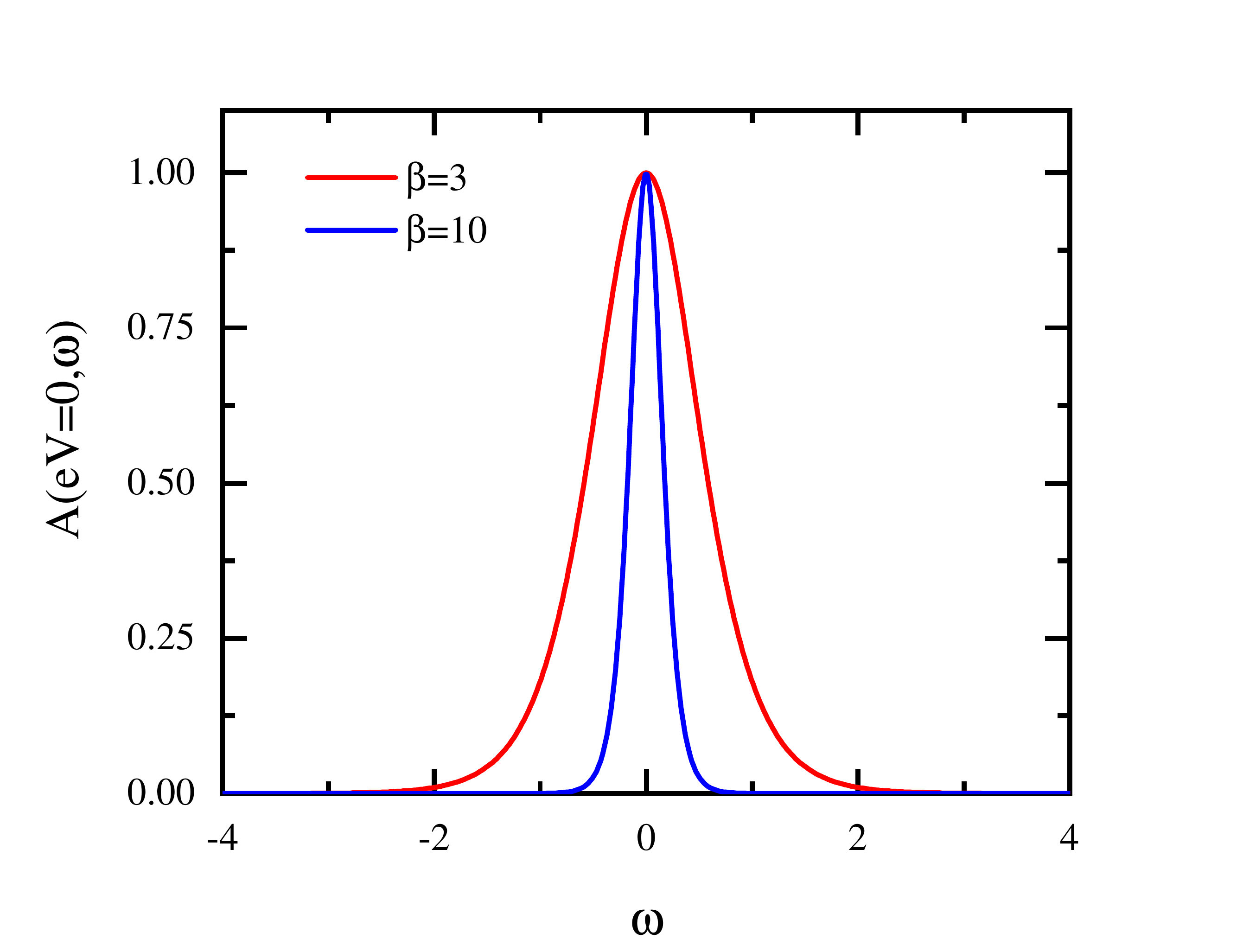}
  \includegraphics[width=0.32\textwidth]{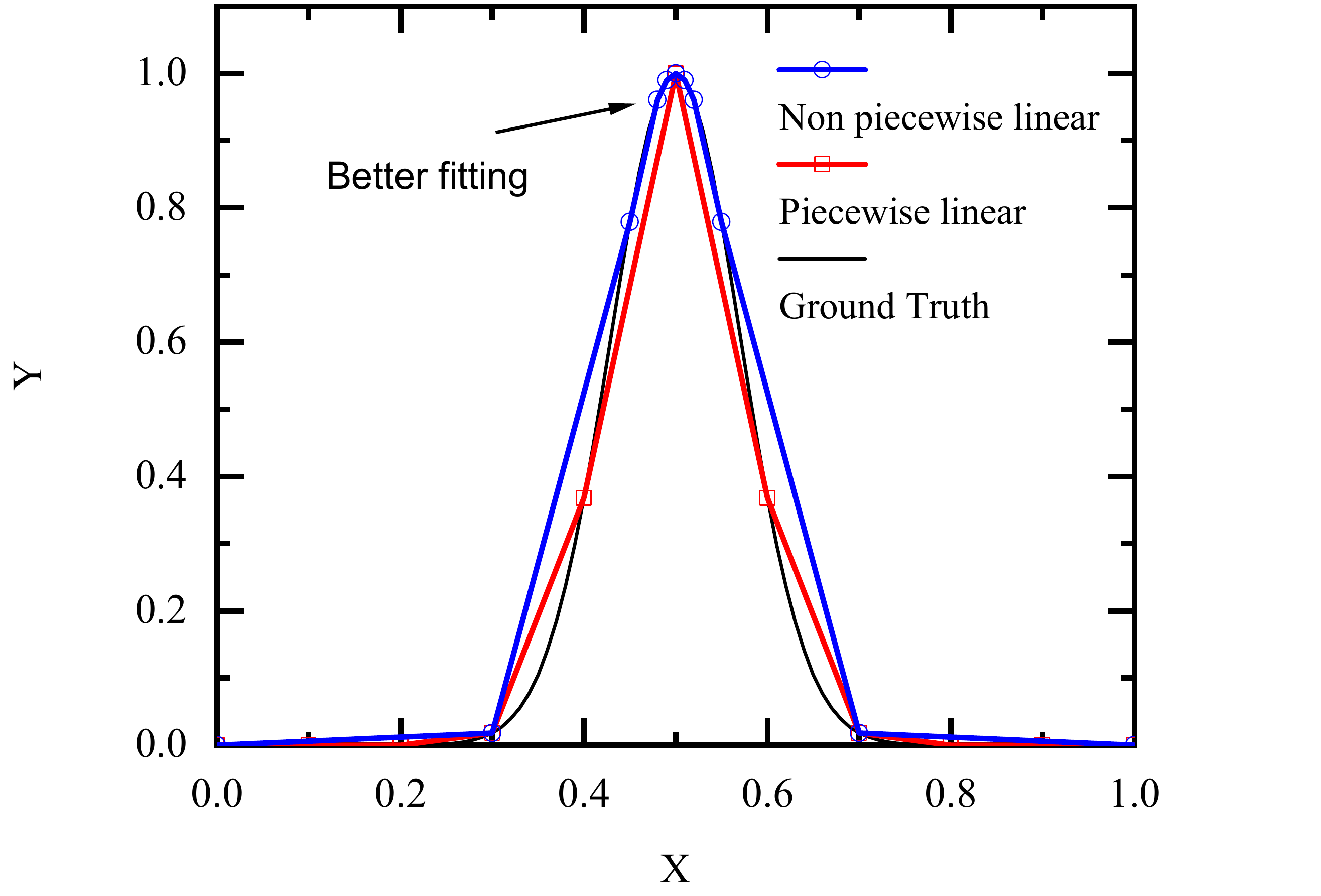}
  \caption{(Upper) The relationship between the condition number of matrix $\bar{A}$ and the sampling rate at different temperatures ($\beta$). The sampling rate is the number of sampling points ($N$) divided by the length of the sampling interval ($\omega$). Subtract 1 from the condition number corresponding to the ordinate and take the logarithm to fully show the characteristics of the curve. (Middle) Broadening properties of $\bar{A}(\epsilon = 0, \omega)$ vary with $\omega$. (Lower) Comparison of fitting curves between piecewise linear function and nonuniform piecewise linear function.}
  \label{fig:cond}
\end{figure}

The condition number of matrix $\bar{A}$ in Eq.\eqref{eq:matrix} is directly related to the temperature and the sampling rate, as in Fig. \ref{fig:cond}(Upper). At the same sampling rate, a lower temperature (larger $\beta$) will effectively reduce the condition number, which can be understood by the broadening properties of $\bar{A}(\epsilon=0,\omega)$ in Fig.\ref{fig:cond}(Middle).
At a fixed temperature, the condition number of the model will diverge with increasing sampling rate at the speed of the transcendental exponential function. This means that a denser sampling rate will bring disastrously unstable results. However, the sampling rate is a direct means to improve the resolution of the results, which undoubtedly leads to a contradiction between resolution and stability.

Therefore, it is very important to select the appropriate sampling rate according to the needs.
MEM and others only consider the case of a uniform mesh; to reduce the loss in Eq.\eqref{eq:loss}, a high sampling rate is needed with a large condition number. This requires the introduction of regularization to avoid oscillation, and the choice of regularization and sampling rate is significant and complicated.
We consider choosing an appropriate high sampling rate for key areas and a low sampling rate for other areas using the nonuniform piecewise linear method. Although it does not significantly improve the condition number, which depends more on regions with dense sampling, it has a significant effect on improving the accuracy of the solution.
SOM can use a nonuniform mesh and can adaptively solve a good mesh. However, it is still necessary to achieve a given strategy of meshes in the calculation process, which is usually difficult.
In this way, we believe that the possible improvement direction is to pay different attention to different intervals.

\subsection{Nonuniform piecewise linearity}

The effect of fitting curves of piecewise linear function and nonuniform piecewise linear function is compared in Figure~\ref{fig:cond}(Lower). The two curves fix $11$ discrete points. The piecewise linear function is discretized uniformly according to $0.1$, which is obviously in the area where the curve changes sharply in the middle, and the curve fitting error is large. A nonuniform piecewise linear function concentrates more discrete points in the interval where the function changes sharply, so it can fit the curve of the middle region more accurately. It needs to be clarified that when evaluating the curve fitting effect, the mean square error of uniform sampling is still used for evaluation, and the evaluation of uniform sampling of mean square error will not affect the fitting effect of piecewise linear and nonuniform piecewise linear functions. 

\subsection{Neural network method}

Therefore. The building of a proper piecewise mesh is a problem that needs further exploration.
In this regard, traditional methods may have difficulty achieving ideal results, but a method based on deep learning may make a breakthrough.
Reference~\cite{xu2017} used neural networks to solve inverse problems, and the core idea was to use the strong expression ability of the neural network itself.
Further theoretical work on neural networks shows that neural networks more easily fit low-frequency information~\cite{2020Frequency,luo2019theory,ma2021frequency}. This phenomenon is called the F-Principle, which implies an implicit bias that neural networks tend to fit training data from a low-frequency function and provides an explanation of the good generalizations of neural networks on most real datasets and the bad generalizations of neural networks on parity functions or randomized datasets. Considering the oscillation problem, which has not been solved by previous methods, this F-principle may make considerable improvements. 

We find that a fully connected neural network with the ReLU activation function can represent piecewise linear functions. The training process of a neural network is equivalent to training a node of a piecewise linear function. These nodes are obviously nonuniform, which can meet our need to find the best nonuniform piecewise mesh.
We introduce a neural network method into the ill-posed problem examined in this paper, named the neural network replacing spectrum (NNRS) method.
The core idea is to replace function $\rho(\omega)$ with a neural network and hope that when we obtain the trained neural network, we can use its inference procedure to give $\rho(\omega)$ corresponding to $\omega$.
\begin{eqnarray}
{\omega} \rightarrow \text{NNRS} \rightarrow \vec{\rho}(\omega).
\end{eqnarray}
Considering the characteristics of $\rho(\omega)$, we adopt a multilayer fully connected neural network with the ReLU activation function, which is mathematically equivalent to a piecewise linear function. It should be noted that we introduce the ReLU activation function in the output layer to ensure that the output is greater than $0$ according to the physical principle of the DOS.

\begin{algorithm}[t]
\caption{Neural Network Replacing Spectrum (NNRS) Method}
\label{alg:train}
	\begin{algorithmic}
		\Require {Experimental data $\{\epsilon,\vec{\sigma}(\epsilon)\}$, hyperparameters~(learning rate $\alpha$, and epochs $n$).}
		\Ensure {Fully connected neural network with parameters $\theta$ in $L$ layers $m$ neurons, corresponding to $\vec{\rho}_{\text{NN}}(\omega)$.}
		\State {Initialize model parameters $\theta$ and sampling ${\omega}$.}
		\\
		\Repeat{$\mathcal{L}$ converge or reach $n$ epochs.}
		{
		\State {Obtain $\vec{\rho}_{\text{NN}}(\omega)$ from $\omega$,} \Comment{Inference of model}
		\State {Obtain $\bar{A}({\epsilon},{\omega})$,} \Comment{Calculating matrix elements}
		\State {$\vec{\sigma}_{\text{NN}}(\epsilon) = \bar{A}(\epsilon,\omega) \vec{\rho}_{\text{NN}}(\omega)$,} \Comment{Calculate Eq.\eqref{eq:matrix}}
		\State {$\mathcal{L} = || \vec{\sigma}(\epsilon) - \vec{\sigma}_{\text{NN}}(\epsilon)  ||_2^2 $, } \Comment{Calculate loss}
		\State {$\theta = \theta - \alpha \nabla_{\theta} \mathcal{L}$,} 
		    \Comment{Update the model with loss}
	    } 
	\end{algorithmic}
\end{algorithm}

After defining the network structure, the flow of the algorithm is shown in Algorithm~\ref{alg:train}. The algorithm starts with the experimental dataset $\{\epsilon,\sigma(\epsilon)\}$, and the first step is to sample $\omega$.
We formally use the uniform discretization of the $[\omega_{min},\omega_{max}]$ interval to sample $N_\omega$ $\omega$, where $N_\omega = N_{\epsilon}$. We set $\omega_{max/min}=\pm 6$ and $20\%$ to expand by approximately $\epsilon_{max/min}$ to ensure that the process of $\epsilon$ broadening is not affected by $\omega$ at the boundary.
Note that sampling $\omega$ is not equal to the nonuniform piecewise node.
NNRS can adopt more flexible sampling methods, including nonuniform sampling and even a different sampling in each iteration step. 

Next, go to the loop iteration. First, $\vec{\rho}_{NN}(\omega)$ is obtained by a forward inference of the neural network. Then, matrix $\bar{A}(\epsilon,\omega)$ is combined to calculate $\vec{\sigma}_{\text{NN}}(\epsilon) = \bar{A}(\epsilon,\omega) \vec{\rho}_{\text{NN}}(\omega)$ as Eq.\eqref{eq:matrix}. At the end of each loop, the derivative of the model parameters $\theta$ is calculated according to the loss function $\mathcal{L} = || \vec{\sigma}(\epsilon) - \vec{\sigma}_{\text{NN}}(\epsilon)  ||_2^2$, and then the model parameters are updated. Here, we use the widely used minimum square error with the $\ell_2$ norm, and generally, we can choose a more appropriate error formula according to the characteristics of the problem.
We use the Adam optimizer with a learning rate of $0.001$ during training.
Finally, the residual error between $\vec{\rho}_{\text{NN}}(\omega)$ and the ground truth $\vec{\rho}(\omega)$ is evaluated.

\subsection{Accuracy metric}

The characterization of the application efficiency of this method and the optimization loss function of the neural network need to be described in detail here. The loss function used to optimize the neural network is described in Algorithm~\ref{alg:train}, and the mean square error is used to ensure that the approximation error is as small as possible, which is consistent with the existing methods.

In the process of evaluating the application efficiency, we not only use the residual error between groundtruth $\{\omega,\vec{\rho}(\omega)\}$, $\text{Err} = || \vec{\rho}(\omega) - \vec{\rho}_{\text{NN}}(\omega) ||_2^2$ as the existing methods but also consider using the position of the peak, the degree of oscillation and whether it is nonnegative as an evaluation index.

\section{Experiments}

\subsection{Theoretical datasets}

\begin{figure}
  \centering
  \includegraphics[width=0.48\textwidth]{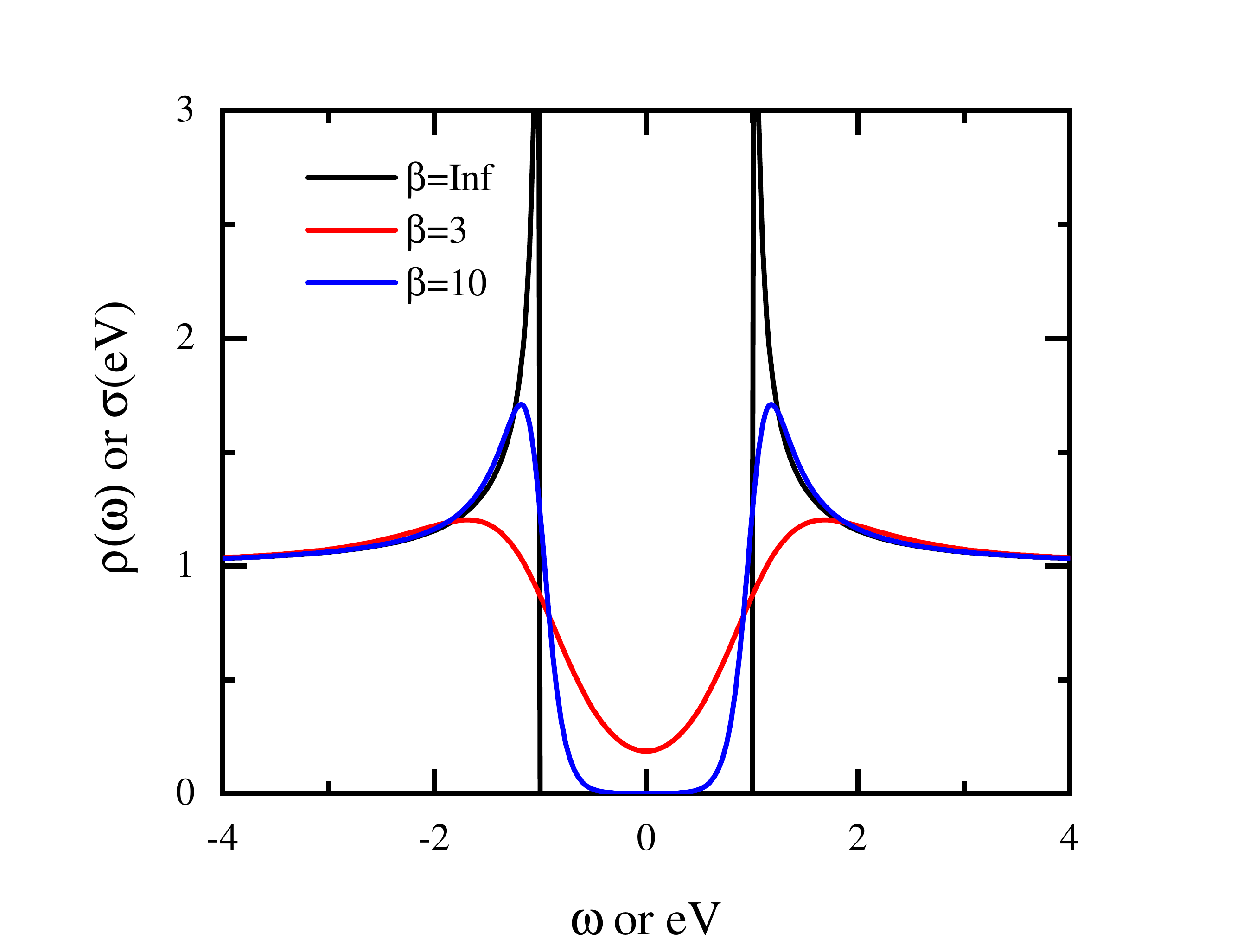}
  \caption{The theoretical curves of the objective function $\rho(\omega)$ and experimental data $\sigma(\epsilon)$ are compared. The black line with temperature $\beta = \text{Inf}$ corresponds to objective function $\rho(\omega)$ with divergence at $\omega = \pm \Delta$, and the red and blue lines with $\beta = 3, 10$ correspond to experimental data $\sigma(\epsilon)$ with broadening.}
  \label{fig:dataset}
\end{figure}

\begin{figure*}
  \centering
  \includegraphics[width=0.48\textwidth]{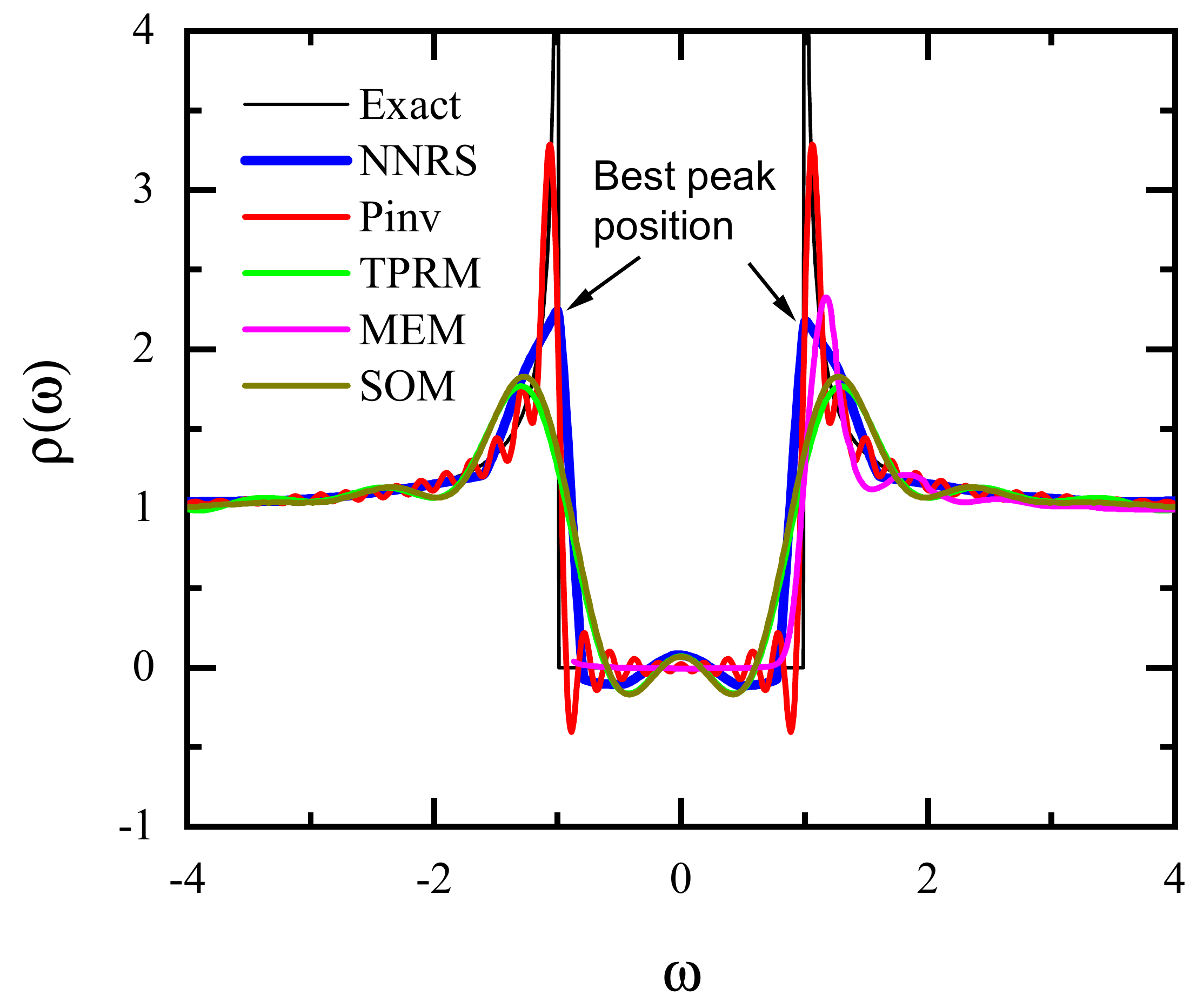}
  \includegraphics[width=0.48\textwidth]{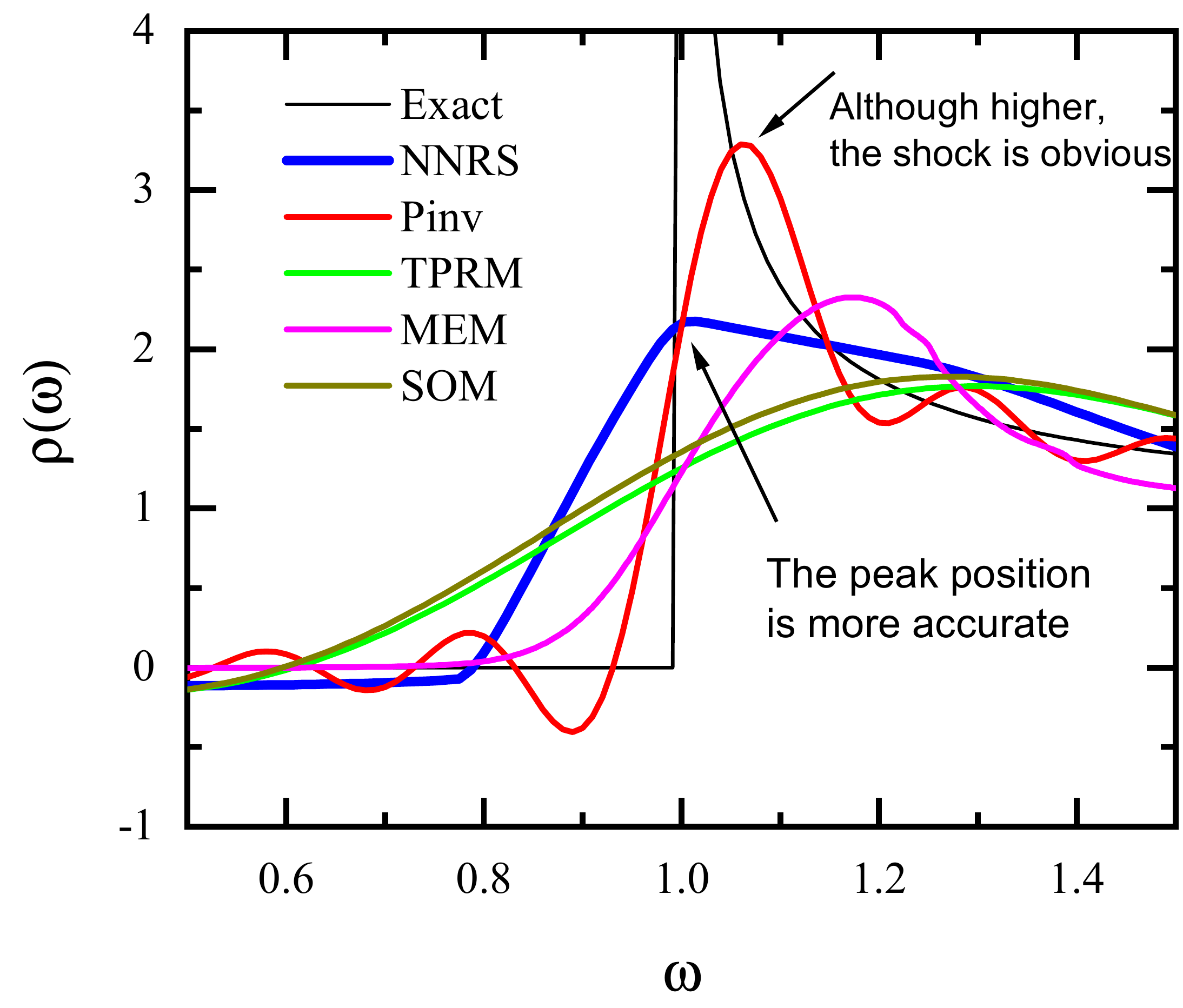}
  \\
  \includegraphics[width=0.48\textwidth]{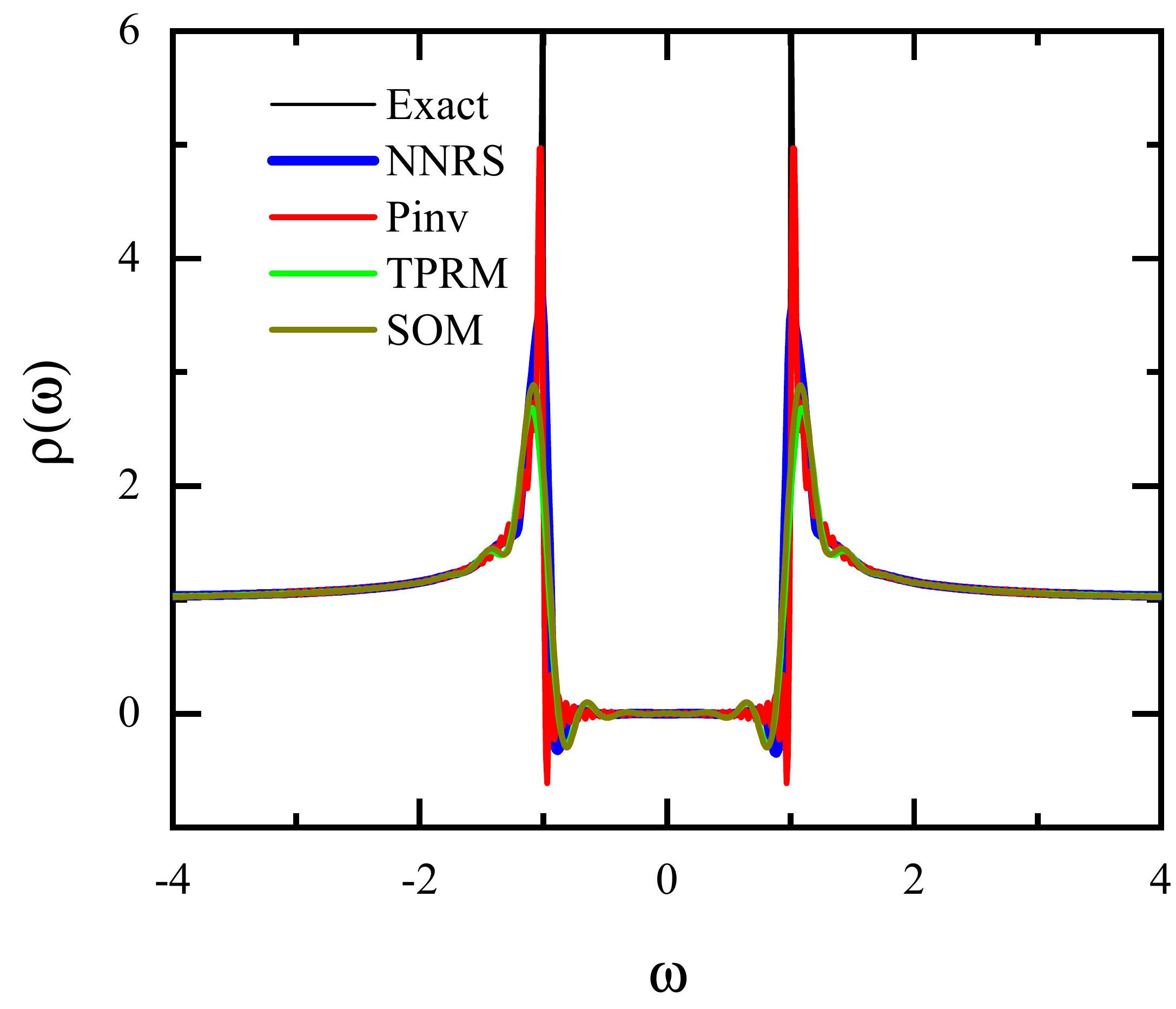}
  \includegraphics[width=0.48\textwidth]{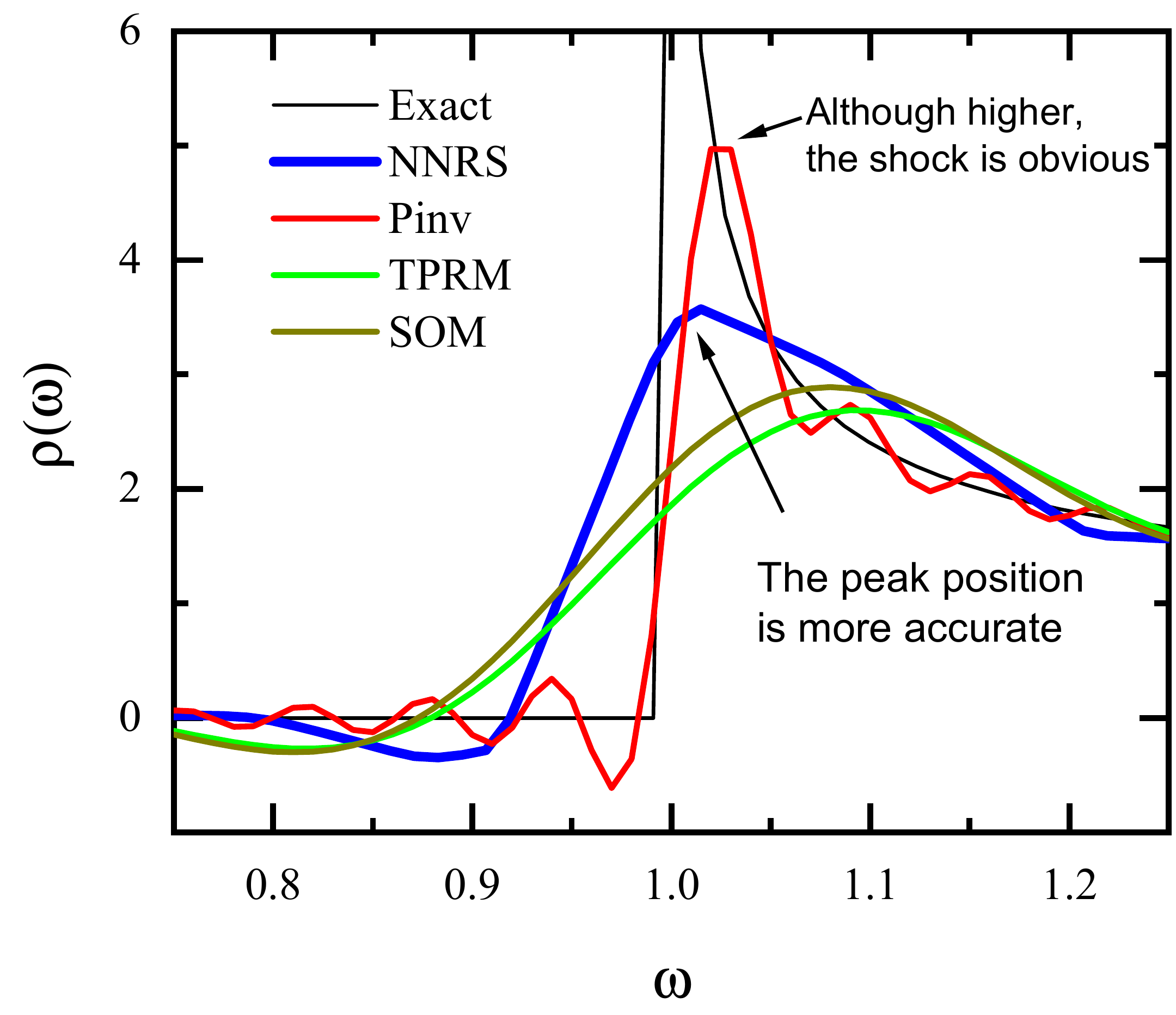}
  \caption{Results of exact dataset with different temperatures of several algorithms as shown in the figure. The corresponding temperatures of the upper and lower lines are $\beta=3$ and $\beta=10$, respectively. The $2$ column images in each row are the deconvolution result panorama, enlarged view around $\Delta$. The neural network used in this figure is a fully connected neural network with $3$ layers and $1000$ neurons in each layer with the ReLU activation function. NNRS performs with the best peak position, while Pinv outputs the highest peak. }
  \label{fig:result1}
\end{figure*}

We use the theoretical formula of isotropic s-wave superconductivity to build the datasets. We can evaluate the model by comparing the solution error of the DOS with that of a strict solution. The exact solution of DOS is
\begin{eqnarray}
\rho\left(\omega\right) & = & \frac{N_{F}\omega \theta(|\omega|-\Delta)}{\sqrt{\omega^{2}-\Delta^{2}}},
\end{eqnarray}
where $N_F$ is the density of states on the Fermi surface, $\Delta$ is the superconducting gap and $\theta(|\omega|-\Delta) = 1$ if $|\omega|>\Delta$, $\theta(|\omega|-\Delta) = 0$ otherwise.
In the process of building a dataset, for the convenience of calculation, we take the following parameters: $N_F = 1$, $\Delta = 1$ and $a = 1$ (in Eq.\eqref{eq:matrix}). 

Figure~\ref{fig:dataset} shows the curves of DOS $\rho(\omega)$ and differential conductance $\sigma(\epsilon)$ with $\beta = 3$ and $10$. It is clear that the zero-temperature DOS diverges at $\omega=\pm \Delta$, and the broadening of $\sigma(\epsilon)$ increases with increasing temperature.
The gap in the middle becomes smooth, the gradient becomes slow, and the divergence at $\omega = \pm \Delta$ disappears. Therefore, the dataset contains $\{\epsilon,\sigma(\epsilon)\}$ for training and $\{\omega,\rho(\omega)\}$ for calculating the approximation error of the evaluation of the algorithms.
$\epsilon_{\min}$, $\epsilon_{\max}$, $N_{\epsilon}$, $\omega_{\min}$, $\omega_{\max}$ and $N_{\omega}$ represent the upper and lower limits of variables $\epsilon$, $\omega$ and the number of samples, respectively.
We set $\epsilon_{\max/\min}=\pm 5$, $\omega_{\max/\min}=\pm 6$ and $N_{\epsilon/\omega} = 1201$ if not specified.

\subsection{Numerical results}

First, we use theoretical datasets to compare the residuals of different approximation functions, as shown in Fig. \ref{fig:result1}. For a low temperature ($\beta = 10$), it is easier to obtain more accurate results, and all the methods show that the result is closer to the exact solution for a low temperature. 
For the interval with gentle curve change, all methods have good performance, because the ill-conditioned problem solving is not significant at this time.
The Pinv method shows the highest peak at approximately $\pm \Delta$ but with the most severe oscillation. Due to the introduction of regularization constraints, the peaks of the TPRM, MaxEnt, and SOM methods are not as high as those of Pinv and exhibit a weaker oscillation. 

The NNRS has good performance in the height and position of the peak and amplitudes of oscillation.
The peak position solved by the NNRS method is the most accurate, reaching $1.01$ and $1.015$, respectively, under the conditions of $\beta=3$ and $\beta=10$, while the peak position obtained by the closest SOM method is $1.06$ and $1.025$. Compared with the strict solution $1$, the deviation of our method is much reduced, and the curve oscillation at the gap is obviously the smallest.
The core of NNRS is using a neural network to approximate the actual function $\rho(\omega)$. 
See Table~\ref{tab} for all peak positions and heights within the experiments. 
Compared with the other methods, it has many advantages. 
\begin{enumerate} 
	\item Because the neural network is a piecewise linear function, mathematically, it can find the appropriate piecewise method independently based on its ability to approach the objective function. Therefore, the neural network can encrypt the piecewise density in the changing interval according to the details of the objective function and reduce the piecewise density in other intervals to achieve a better approximation mesh by using the finite number of segments effect.

	\item The obtained neural network does not depend on the sample points used in training and has considerable generalization abilities; that is, it can give all the output results corresponding to the input in the specified interval.

	\item The ReLU activation function can ensure that the model is always greater than $0$; it is strictly positive.
\end{enumerate}

\begin{table}[]
\caption{Comparison of peak positions and heights of different methods in different experimental results.}
\label{tab}
\centering
\footnotesize
\begin{tabular}{ccccccc}
\toprule
\multicolumn{2}{l}{Data} & NNRS & Pinv & TPRM & MEM  & SOM  \\
\midrule
\multirow{2}{*}{Theoretical dataset $\beta = 3$}   & $\omega$ & \textbf{1.02} & 1.06 & 1.30 & 1.17 & 1.27 \\
                     & $\rho(\omega)$ & 2.18 & 3.29 & 1.77 & 2.33 & 1.83 \\
                     \midrule
\multirow{2}{*}{Theoretical dataset $\beta = 3$}  & $\omega$ & \textbf{1.02} & 1.03 & 1.09 & 1.08 & 1.08 \\
                     & $\rho(\omega)$ & 3.57 & 4.97 & 2.69 & 2.90 & 2.89 \\
                     \midrule
\multirow{2}{*}{experimental dataset~\cite{Ge2017Anisotropic}}   & $\omega$ & \textbf{2.23} & inf  & 2.43 & 2.53 & 2.43 \\
                     & $\rho(\omega)$ & \textbf{1.31} & inf  & 1.22 & 1.23 & 1.03 \\
\bottomrule
\end{tabular}
\end{table}


\subsection{Network hyperparameters}

Because the training process of the neural network requires a large number of hyperparameters, the influence of these parameters is briefly discussed here. Overall, our method has little dependence on parameters.

First, in terms of network structure, this method uses a fully connected neural network, mainly including a number of neural network layers and a number of neurons in each layer. Obviously, when the number of neurons is too small ($<O(10)$), the degree of freedom of the neural network is too low, which is equivalent to the discrete sampling points being too sparse. When this happens, the segmentation characteristics of the training results are obvious. As long as the network complexity is increased, if either the number of layers or the number of neurons is increased, the training results can be effectively improved. Most of the examples in this experiment use a $3$-layer fully connected network with $1000$ neurons in each layer.
If the data scale increases, the scale of the neural network needs to increase accordingly. However the overall quality of the results is not sensitive to the fine adjustment of the neural network. 

The NNRS method only depends on $1$ sets of data and does not need a large-scale dataset composed of thousands of sets of data, as in the field of traditional computer vision.
Therefore, for a model's architecture and iteration times that are not exaggerated, the calculation results are sufficient to meet the needs, and the training time is still short.

In the training process, the learning rate and the number of epochs are usually a pair of key and interactive parameters.
First, to converge quickly, starting from the randomly initialized neural network at the beginning of training, an appropriately larger learning rate should be used to speed up the network update and evolution. When the network learns an approximate result, it is necessary to gradually reduce the learning rate to enable the network to learn with precision and further converge. In practical experience, the learning rate of $0.01$ can be used for the first $100$ epochs, and then the learning rate of $0.001$ can be used to continue training for $500$ epochs. If you want to obtain more refined results, you can further reduce the learning rate to $0.0001$ for another $1000$ epochs.
At this time, the training loss will converge to a high and stable level. It is generally believed that the loss will converge when it reaches $10^{-6}$.

Finally, it should be noted that although the result of this method is robust in most cases, it still depends on the initial value of the neural network. That is, if the random initial value is insufficient, the training result is likely to be unsatisfactory. Therefore, the training is usually repeated $2-3$ times, and the best or most stable result is taken as the final result. Considering that each training time is a few minutes, the time cost caused by multiple training sessions is completely acceptable.

\begin{figure}
  \centering
  \includegraphics[width=0.48\textwidth]{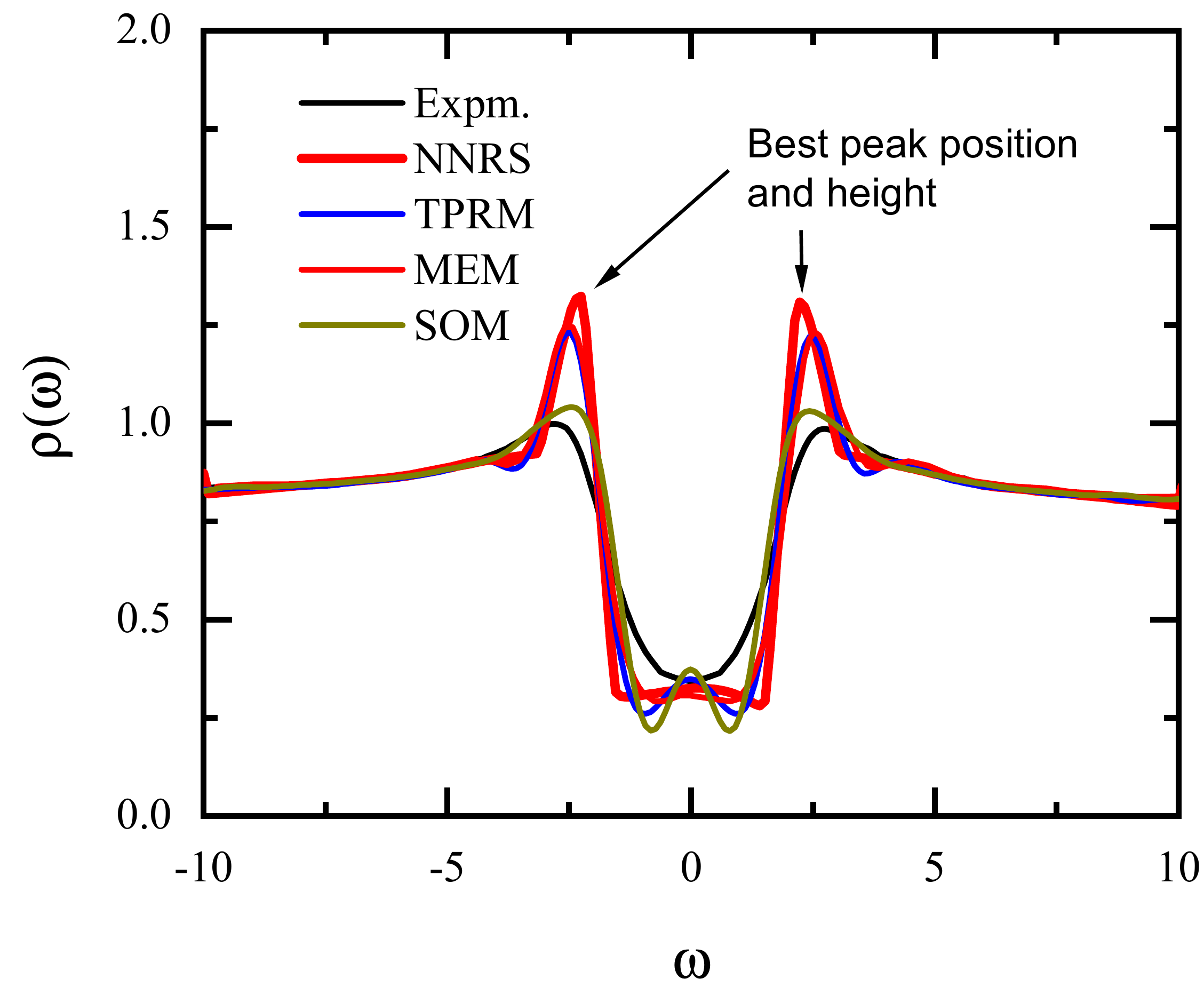}
  \includegraphics[width=0.48\textwidth]{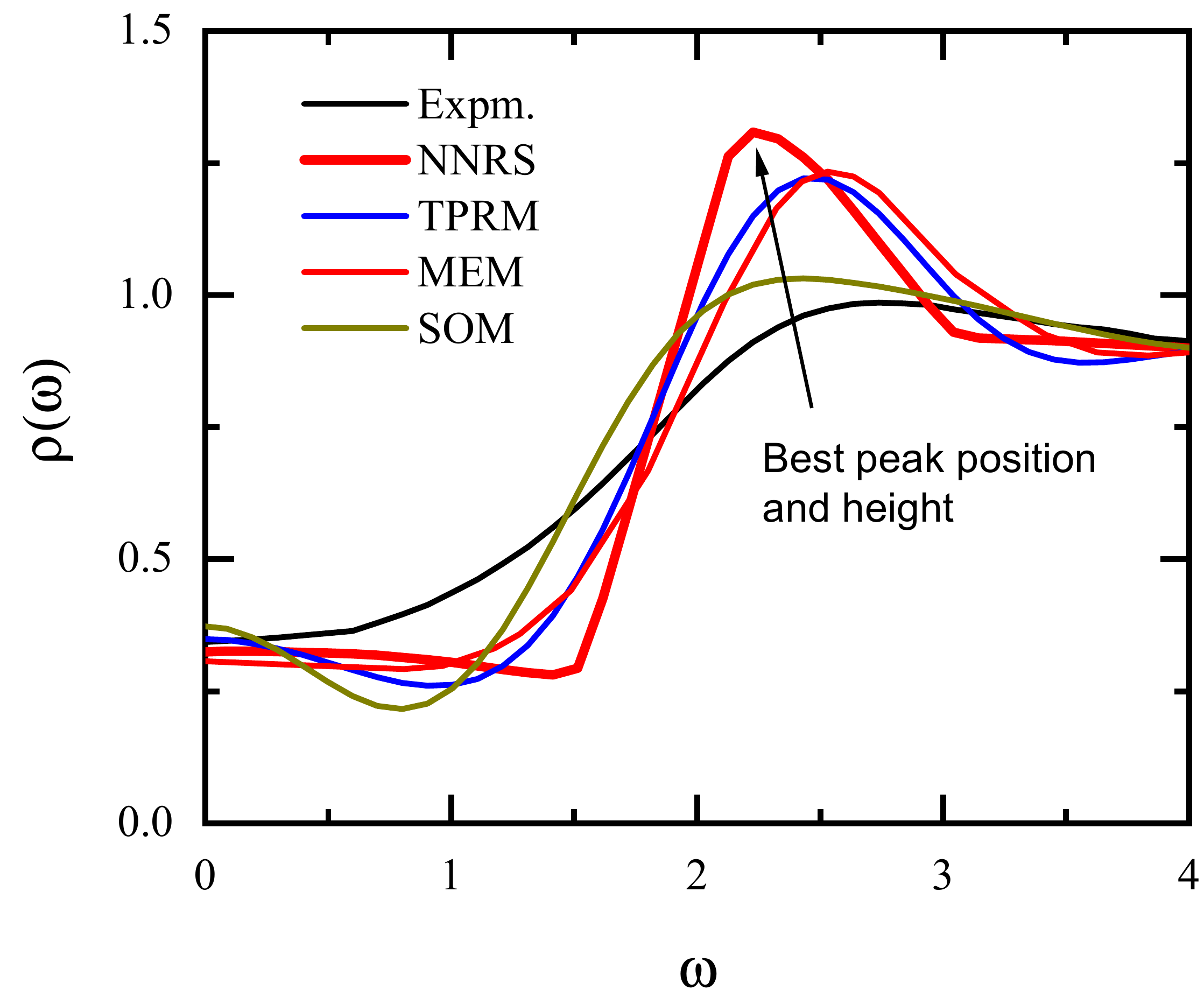}
  \caption{Results of experimental dataset~\cite{Ge2017Anisotropic} with temperature $\beta =  2.32$ of several algorithms as shown in the figure (Upper) full scale and (Lower) amplification. The neural network used in this figure is a fully connected neural network with $3$ layers and $1000$ neurons in each layer without an activation function on the output layer. NNRS performs with the best peak and high position, while Pinv diverges and is not drawn on the figure. }
  \label{fig:resultExp}
\end{figure}


\subsection{Real experimental data}

The previous results from the theoretical curve can show the convergence of the algorithm, but the actual experimental data usually have a certain degree of noise error. This subsection observes the robustness of these methods to the error data through the actual experimental data.

Figure~\ref{fig:resultExp} shows the deconvolution results of several algorithms for real experimental data~\cite{Ge2017Anisotropic} (the differential conductivity at $T = 2.5K$ without a magnetic field at the crystal surface measured by STM). The biggest difference between experimental data and theoretical data is that they contain nonnegligible random noise, so the Pinv method cannot obtain stable results and will not be shown. In comparison, the regularization constraints introduced by the TPRM and MEM play a key role in noise suppression, and smooth results are obtained. The SOM method mainly combines TPRM results with different parameters, and the result is not outstanding compared to TPRM.

It can be seen from the figure that the curve peak obtained by the NNRS method is the clearest and can be used to estimate the superconducting energy gap. 
The peak position and height obtained by the NNRS method are the best. The peak position at the right end is $2.23$, which is nearly doubled compared with the most competitive peak position of $2.43$. See Table~\ref{tab} for all peak positions and heights within the experiments. 
This fully reflects the advantages of the NNRS method in solving this problem and the mining of information. Using the characteristics of the neural network's first low-frequency information, we can obtain more accurate and nonoscillatory results. 

\subsection{Potential shortcomings}
At the end of the discussion, the author indicates that this method still has some potential shortcomings, but the disadvantages of this method as a whole do not attract people's attention. Specifically, (1) calculation amount: compared with traditional MEM and other methods, the calculation amount of neural network operation in this method has no additional advantages. Although the overall calculation time is only a few seconds, it is difficult to further improve the time cost at present. (2) Mechanism: This method uses a neural network. Compared with MEM and other optimization methods, the academic community generally believes that the theoretical mechanism behind the neural network solution algorithm is not clear. (3) The imaginary number, MEM and other methods are based on mathematical formulas and can be used for physical problems containing imaginary numbers. This method uses neural networks, which cannot be proven to perform well in problems containing imaginary numbers.

\section{Conclusion}

Physical experiments or numerical methods cannot obtain target physical dynamical properties directly. It must use the available data to deconvolute the dynamical information, such as to obtain the density of states from differential conductance in scanning tunneling microscopy. This problem is essentially ill-posed and is unstable due to sawtooth noise in the input data. The existing methods can obtain stable results but rely on prior knowledge and are unsatisfactory.
In this way, we propose the neural network replacing spectrum (NNRS) method for this problem by using a fully connected neural network to approach the objective function $\rho(\omega)$ and training the network using experimental observation data. The results of our method are smooth and stable, and the position of the approaching peak is closer to the exact solution. After using the output layer ReLU activation function and other tricks, the results are obviously of high value.

In terms of practicability, our method is an out of the box method similar to MEM, which does not need the dataset construction and complex training process of the general deep learning model, nor does it need experienced parameter adjustment. To facilitate follow-up research, we plan to expose the open source code of the method to GitHub, ``https://github.com/erickxhd/NNRS.git''. At the same time, the practicability of this method has also been recognized by many experimental physics research groups, and many related works have tried to use this method.

In terms of expansibility, our approach can be extended to many other similar problems in physics, such as experimental areas such as obtaining Lehmann functions from the spectral response observed in experiments on angle-resolved photoelectron spectroscopy (ARPES)~\cite{RevModPhys.75.473} and numerical areas such as analytic continuation problems pertaining to obtaining dynamic correlation functions from quantum Monte Carlo simulations with a finite lattice size and imaginary time~\cite{PhysRevB.61.9300}.
Moreover, similar equations have to be solved for medical X-ray and impedance tomography, image deblurring, and many other practical applications~\cite{Kaipio2015Statistical}.

\section*{Acknowledgments}

This work was supported by the National Natural Science Foundation of China (Grant No. 12004422) and by Beijing Nova Program of Science and Technology (Grant No. Z191100001119129). We thank researcher Tao Xiang and researcher Lei Wang of the Institute of Physics, Chinese Academy of Sciences, for their guidance and help in the completion of this work.

\section*{References}

\bibliographystyle{unsrt}
\bibliography{ref}

\begin{thebibliography}{10}

\bibitem{1983Condensed}
J.~E. Hirsch and D.~J. Scalapino.
\newblock Condensed-matter physics.
\newblock {\em Physics Today}, 36(5):44--52, 1983.

\bibitem{1998Metal}
Masatoshi Imada, Atsushi Fujimori, and Yoshinori Tokura.
\newblock Metal-insulator transitions.
\newblock {\em Rev. Mod. Phys.}, 70:1039--1263, Oct 1998.

\bibitem{1993Correlated}
Elbio Dagotto.
\newblock Correlated electrons in high-temperature superconductors.
\newblock {\em Rev. Mod. Phys.}, 66:763--840, Jul 1994.

\bibitem{Park1987Scanning}
Sang-Il {Park} and C.~F. {Quate}.
\newblock {Scanning tunneling microscope}.
\newblock {\em Review of Scientific Instruments}, 58(11):2010--2017, November
  1987.

\bibitem{PhysRevLett.6.57}
J.~Bardeen.
\newblock Tunnelling from a many-particle point of view.
\newblock {\em Phys. Rev. Lett.}, 6:57--59, Jan 1961.

\bibitem{Teodorescu2013Mathematical}
A.~G. Ramm.
\newblock {\em Mathematical and Analytical Techniques with Applications to
  Engineering}.
\newblock Springer Science \& Business Media, 2006.

\bibitem{0Inverse}
Sergey~I. Kabanikhin.
\newblock {\em Inverse and Ill-posed Problems: Theory and Applications}.
\newblock De Gruyter, 2011.

\bibitem{Linear1999Rainer}
Rainer Kress.
\newblock {\em Linear Integral Equations}.
\newblock Springer, New York, NY, 1999.

\bibitem{Jarrell2012MEM}
Mark Jarrell.
\newblock The maximum entropy method: Analytic continuation of qmc data.
\newblock {\em Correlated Electrons: From Models to Materials, Modeling and
  Simulation}, 2(2), 2012.

\bibitem{Mishchenko2012Stochastic}
Andrey~S Mishchenko.
\newblock Stochastic optimization method for analytic continuation.
\newblock {\em Correlated Electrons: From Models to Materials, Modeling and
  Simulation}, 2(2), 2012.

\bibitem{Golub1965Calculating}
G.~Golub and W.~Kahan.
\newblock Calculating the singular values and pseudo-inverse of a matrix.
\newblock {\em Journal of the Society for Industrial and Applied Mathematics
  Series B Numerical Analysis}, 2(2):205--224, 1965.

\bibitem{Tikhonoff1943}
A.N. Tikhonoff.
\newblock On the stability of inverse problems.
\newblock {\em Dokladyu Akademii Nauk SSSR}, 39:195, 1943.

\bibitem{Tikhonoff1963}
A.N. Tikhonoff.
\newblock Resolution of ill-posed problems and the regularization method.
\newblock {\em Dokladyu Akademii Nauk SSSR}, 151:501, 1963.

\bibitem{Phillips1962}
D.L. Phillips.
\newblock A technique for the numerical solution of certain integral equations
  of the first kind.
\newblock {\em J. ACM}, 9:84, 1962.

\bibitem{JARRELL1996133}
Mark Jarrell and J.E. Gubernatis.
\newblock Bayesian inference and the analytic continuation of imaginary-time
  quantum monte carlo data.
\newblock {\em Physics Reports}, 269(3):133--195, 1996.

\bibitem{Vafayi2007Analytical}
K.~Vafayi and O.~Gunnarsson.
\newblock Analytical continuation of spectral data from imaginary time axis to
  real frequency axis using statistical sampling.
\newblock {\em Phys. Rev. B}, 76:035115, Jul 2007.

\bibitem{Fournier_2020}
Romain Fournier, Lei Wang, Oleg~V. Yazyev, and QuanSheng Wu.
\newblock Artificial neural network approach to the analytic continuation
  problem.
\newblock {\em Phys. Rev. Lett.}, 124:056401, Feb 2020.

\bibitem{Yoon_2018}
Hongkee Yoon, Jae-Hoon Sim, and Myung~Joon Han.
\newblock Analytic continuation via domain knowledge free machine learning.
\newblock {\em Phys. Rev. B}, 98:245101, Dec 2018.

\bibitem{Arsenault_2017}
Louis-Fran{\c{c}}ois Arsenault, Richard Neuberg, Lauren~A Hannah, and Andrew~J
  Millis.
\newblock Projected regression method for solving fredholm integral equations
  arising in the analytic continuation problem of quantum physics.
\newblock {\em Inverse Problems}, 33(11):115007, oct 2017.

\bibitem{Skilling2002Probabilistic}
J.~Skilling.
\newblock Probabilistic data analysis: an introductory guide.
\newblock {\em Journal of Microscopy}, 190(1-2):28--36, 2002.

\bibitem{Sandvik1998Stochastic}
Anders~W. {Sandvik}.
\newblock {Stochastic method for analytic continuation of quantum Monte Carlo
  data}.
\newblock {\em Phys. Rev. B}, 57(17):10287--10290, May 1998.

\bibitem{Mishchenko2000Diagrammatic}
A.~S. Mishchenko, N.~V. Prokof'ev, A.~Sakamoto, and B.~V. Svistunov.
\newblock Diagrammatic quantum monte carlo study of the fr\"ohlich polaron.
\newblock {\em Phys. Rev. B}, 62:6317--6336, Sep 2000.

\bibitem{Cybenko1989Approximation}
G.~Cybenko.
\newblock Approximation by superpositions of a sigmoidal function.
\newblock {\em Math. Control Signal Systems}, 2(303-314), 1989.

\bibitem{Xu2018Understanding}
Zhiqin~John Xu.
\newblock Understanding training and generalization in deep learning by fourier
  analysis.
\newblock 2018.

\bibitem{2020SciPy-NMeth}
Pauli Virtanen, Ralf Gommers, Travis~E. Oliphant, Matt Haberland, Tyler Reddy,
  David Cournapeau, Evgeni Burovski, Pearu Peterson, Warren Weckesser, Jonathan
  Bright, St{\'e}fan~J. {van der Walt}, Matthew Brett, Joshua Wilson, K.~Jarrod
  Millman, Nikolay Mayorov, Andrew R.~J. Nelson, Eric Jones, Robert Kern, Eric
  Larson, C~J Carey, {\.I}lhan Polat, Yu~Feng, Eric~W. Moore, Jake
  {VanderPlas}, Denis Laxalde, Josef Perktold, Robert Cimrman, Ian Henriksen,
  E.~A. Quintero, Charles~R. Harris, Anne~M. Archibald, Ant{\^o}nio~H. Ribeiro,
  Fabian Pedregosa, Paul {van Mulbregt}, and {SciPy 1.0 Contributors}.
\newblock {{SciPy} 1.0: Fundamental Algorithms for Scientific Computing in
  Python}.
\newblock {\em Nature Methods}, 17:261--272, 2020.

\bibitem{PhysRevB.44.6011}
J.~E. Gubernatis, Mark Jarrell, R.~N. Silver, and D.~S. Sivia.
\newblock Quantum monte carlo simulations and maximum entropy: Dynamics from
  imaginary-time data.
\newblock {\em Phys. Rev. B}, 44:6011--6029, Sep 1991.

\bibitem{PhysRevB.61.5147}
K.~S.~D. Beach, R.~J. Gooding, and F.~Marsiglio.
\newblock Reliable pad\'e analytical continuation method based on a
  high-accuracy symbolic computation algorithm.
\newblock {\em Phys. Rev. B}, 61:5147--5157, Feb 2000.

\bibitem{1989Classic}
J.~Skilling.
\newblock {\em Classic Maximum Entropy}, volume~36, chapter~3, pages 45--52.
\newblock Kluwer Academic Publishers, 1989.

\bibitem{PhysRevB.96.155128}
Gernot~J. Kraberger, Robert Triebl, Manuel Zingl, and Markus Aichhorn.
\newblock Maximum entropy formalism for the analytic continuation of
  matrix-valued green's functions.
\newblock {\em Phys. Rev. B}, 96:155128, Oct 2017.

\bibitem{Levy_2017}
Ryan Levy, J.P.F. LeBlanc, and Emanuel Gull.
\newblock Implementation of the maximum entropy method for analytic
  continuation.
\newblock {\em Computer Physics Communications}, 215:149–155, Jun 2017.

\bibitem{keskar2017largebatch}
Nitish~Shirish Keskar, Dheevatsa Mudigere, Jorge Nocedal, Mikhail Smelyanskiy,
  and Ping Tak~Peter Tang.
\newblock On large-batch training for deep learning: Generalization gap and
  sharp minima, 2017.

\bibitem{Krivenko_2019}
Igor Krivenko and Malte Harland.
\newblock Triqs/som: Implementation of the stochastic optimization method for
  analytic continuation.
\newblock {\em Computer Physics Communications}, 239:166–183, Jun 2019.

\bibitem{xu2017}
ZongBen Xu, Yan Yang, and Jian Sun.
\newblock A new approach to solve inverse problems:combination of model-based
  solving and example-based learning.
\newblock {\em SCIENTIA SINICA Mathematica}, 047(010):P.1345--1354, 2017.

\bibitem{2020Frequency}
Zhi-Qin John~Xu, Yaoyu Zhang, Tao Luo, Yanyang Xiao, and Zheng Ma.
\newblock Frequency principle: Fourier analysis sheds light on deep neural
  networks.
\newblock {\em Communications in Computational Physics}, 28(5):1746--1767,
  2020.

\bibitem{luo2019theory}
Tao Luo, Zheng Ma, Zhi-Qin~John Xu, and Yaoyu Zhang.
\newblock Theory of the frequency principle for general deep neural networks,
  2019.

\bibitem{ma2021frequency}
Yuheng Ma, Zhi-Qin~John Xu, and Jiwei Zhang.
\newblock Frequency principle in deep learning beyond gradient-descent-based
  training, 2021.

\bibitem{Ge2017Anisotropic}
Ge~He, Yanli Jia, Xingyuan Hou, Zhongxu Wei, Haidong Xie, Zhenzhong Yang, Jinan
  Shi, Jie Yuan, Lei Shan, Beiyi Zhu, Hong Li, Lin Gu, Kai Liu, Tao Xiang, and
  Kui Jin.
\newblock Anisotropic electron-phonon coupling in the spinel oxide
  superconductor $\mathrm{LiT}{\mathrm{i}}_{2}{\mathrm{o}}_{4}$.
\newblock {\em Phys. Rev. B}, 95:054510, Feb 2017.

\bibitem{RevModPhys.75.473}
Andrea Damascelli, Zahid Hussain, and Zhi-Xun Shen.
\newblock Angle-resolved photoemission studies of the cuprate superconductors.
\newblock {\em Rev. Mod. Phys.}, 75:473--541, Apr 2003.

\bibitem{PhysRevB.61.9300}
Carey Huscroft, Richard Gass, and Mark Jarrell.
\newblock Maximum entropy method of obtaining thermodynamic properties from
  quantum monte carlo simulations.
\newblock {\em Phys. Rev. B}, 61:9300--9306, Apr 2000.

\bibitem{Kaipio2015Statistical}
Jari Kaipio and Erkki Somersalo.
\newblock {\em Statistical and Computational Inverse Problems}.
\newblock Springer, Berlin, 2015.

\end{thebibliography}

\end{document}